\documentclass[iop,apj,tighten,twocolappendix,numberedappendix]{emulateapj}
\usepackage{apjfonts}

\usepackage{graphicx}
\usepackage{amsmath}
\usepackage{amssymb}
\usepackage{color}
\usepackage{mathtools}
\usepackage{url}

\usepackage[breaklinks,colorlinks,citecolor=blue,linkcolor=red]{hyperref} 
\usepackage[all]{hypcap}

\newcommand\msun{\, \rm M_\odot}

\newcommand\kms{\, \rm km\,s^{-1}}

\newcommand\mcrit{{m_{\rm crit}}}

\begin{document}

\title{Binary black hole mergers from young massive and open clusters: comparison to GWTC-2 gravitational wave data}

\author{Giacomo Fragione\altaffilmark{1,2}, Sambaran Banerjee\altaffilmark{3,4}}
\affil{$^1$Department of Physics \& Astronomy, Northwestern University, Evanston, IL 60208, USA} 
\affil{$^2$Center for Interdisciplinary Exploration \& Research in Astrophysics (CIERA)}
\affil{$^3$Helmholtz-Instituts f\"{u}r Strahlen- und Kernphysik (HISKP), Nussallee 14-16, D-53115 Bonn, Germany}
\affil{$^4$Argelander-Institut f\"{u}r Astronomie (AIfA), Auf dem H\"{u}gel 71, D-53121 Bonn, Germany}

\begin{abstract}
Several astrophysical scenarios have been proposed to explain the origin of the population of binary black hole (BBH) mergers detected in gravitational waves (GWs) by the LIGO/Virgo Collaboration. Among them, BBH mergers assembled dynamically in young massive and open clusters have been shown to produce merger rate densities consistent with LIGO/Virgo estimated rates. We use the results of a suite of direct, high-precision $N$-body evolutionary models of young massive and open clusters and build the population of BBH mergers, by accounting for both a cosmologically-motivated model for the formation of young massive and open clusters and the detection probability of LIGO/Virgo. We show that our models produce dynamically-paired BBH mergers that are well consistent with the observed masses, mass ratios, effective spin parameters, and final spins of the second Gravitational Wave Transient Catalog (GWTC-2). 
\end{abstract}


\section{Introduction}
\label{sect:intro}

Together with GWTC-1, from the first two observational runs \citep{lvc2019cat}, the second Gravitational Wave Transient Catalog (GWTC-2) by the LIGO/Virgo Collaboration, from the first half of the third observational run \citep{lvc2020cat}, comprises of $50$ events, which are revolutionizing our understanding of black holes (BHs) and neutron stars (NSs). Thanks to the growing number of detected events, gravitational waves (GWs) provide a unique opportunity to probe fundamental physics and the distributions of masses, spins, and merger rates of stellar remnants can be constrained with unprecedented precision \citep{lvc2020catb,lvc2020catc}.

This extraordinary wealth of data gives an unparalleled opportunity to understand the origin of compact binary mergers. Several astrophysical channels have been proposed, including isolated binary evolution through a common envelope phase \citep{bel16b,gm2018,kruc2018} or through chemically homogeneous evolution \citep{demink2016,march16}, mergers in star clusters \citep{BanerjeeBaumgardt2010,askar17,baner18,frak18,rod18,DiCarloMapelli2020,FragioneSilk2020,kremer2020,MapelliSantoliquido2020,Trani2021}, Kozai-Lidov (KL) mergers of binaries in galactic nuclei \citep{antoper12,petr17,fragrish2018,grish18}, in triple \citep{ant17,sil17,frl2019a,frl2019b,fragetal2020,MichaelyPerets2020} and quadruple systems \citep{fragk2019,liu2019}, mergers in AGN accretion disks \citep{bart17,secunda2019,LiDempsey2021}, and GW capture events in galactic nuclei \citep{olea09,rass2019}.

Most of the scenarios account for roughly the same rate and the statistical contribution of each of them can be disentangled as the number of detected events increases \citep[e.g.,][]{olea16,gondan2018,PernaWang2019,WongBreivik2020,ZevinBavera2020,BouffanaisMapelli2021}. Thus, it is of fundamental importance to identify physical quantities and to provide tools to distinguish among the mergers that originate in different astrophysical channels. It has been shown that useful physical quantities that can help doing so are the masses, spins, eccentricity, and redshift distributions of the merging binaries, which can be used as an indicator to statistically disentangle among the contributions of the several scenarios.

In this paper, we use high-precision self-consistent N-body models of young massive and open clusters to study the properties of the binary BH (BBH) mergers formed dynamically in them and compare to LIGO/Virgo GWTC-2. Our direct $N$-body simulations are performed with the state-of-the-art collisional evolution code \textsc{Nbody7} \citep{aseth2003,aseth2012}, with the most up-to-date prescriptions for single and binary stellar evolution \citep{baner2019bse}. Our suite of $N$-body simulations, presented for the first time in \citet{Banerjee2021}, encompasses different initial cluster masses, fractions of primordial binaries, and metallicities. Additionally, we include several schemes for assigning spins of stellar-remnant BHs based on detailed stellar-evolutionary models, as described in \citet{Banerjee2021}.

Our paper is organized as follows. In Section \ref{sect:models}, we describe our numerical models of dense star clusters. In Section \ref{sect:res}, we discuss the expected mass, mass ratio, and spin distributions of the BBHs that merge in our simulations and compare them to LIGO/Virgo data. Finally, in Section \ref{sect:conc}, we summarize our findings and draw our conclusions.

\section{N-body models of young massive and open clusters}
\label{sect:models}

We utilize a catalog of $65$ direct $N$-body evolutionary models of star clusters, computed using the most up-to-date version of \textsc{Nbody7} \citep{aseth2012}. The updates include the prescriptions for the natal kicks imparted to remnants at formation and the natal BH spins, the inclusions of relativistic recoil kicks as a result of BH-BH mergers, the treatment of stellar winds, and the prescriptions for star-star and star-remnant mergers. For details see \citet{Banerjee2021}.

The cluster models we consider in our analysis are proxy for young massive and open clusters, which continuously form and dissolve throughout gas-rich galaxies, such as in the Milky Way and the Local Group. We model star clusters that have initial sizes $\sim 1$\,pc, consistent with gas-free young clusters in our Galaxy and neighbouring galaxies \citep{portgz2010,banerjkroupa2017}. We assume that these clusters have survived their assembling and violent-relaxation phases, and have expanded to parsec-scale sizes from sub-parsec sizes, as observed in newly-formed, gas-embedded, and partially-embedded clusters and associations \citep[e.g.,][]{bankroupa2018}.

The initial model clusters follow a \citet{plummer1911} profile, with masses, half-mass radii, metallicities in the range $1.0\times 10^4\msun$--$1.0\times 10^5\msun$, $1.0$\,pc--$3.0$\,pc, $0.0001$--$0.02$, respectively. All the simulated models are assumed to be initially in virial equilibrium and unsegregated, subjected to an external solar-neighbourhood-like galactic field \citep[see Table~C1 in][]{Banerjee2021}.

Initial stellar masses are sampled from a canonical initial mass function \citep{kroupa2001}, in the range $0.08\msun$--$150\msun$. The overall primordial binary fraction in our models is set to $0.0$, $0.05$, $0.10$. We separately fix the initial binary fraction of the O-type stars, with ZAMS mass $m_{\rm ZAMS}\ge \mcrit=16\msun$,  to be $\sim 100\,\%$ \citep{baner18}, consistent with the observed high binary fractions among the OB-type stars in young clusters and associations \citep[see, e.g.,][]{sanaevans2011,moedist2017}. For $m_{\rm ZAMS}\ge\mcrit$, we pair an OB-star only with another OB-star, as consistent to observations, and the binaries are taken to initially follow the orbital-period distribution of \citet{sanaevans2011} and a uniform mass-ratio distribution. The pairing among the lower mass stars in primordial binaries is random. Moreover, their orbital periods are sampled to follow the \citet{duq1991} distribution and their mass-ratio distribution is taken to be uniform. The initial binary eccentricities are drawn from a thermal distribution \citep{spitzer1987} for the binaries with components $m_{\rm ZAMS}<\mcrit$ and from the \citet{sanaevans2011} distribution for the $m_{\rm ZAMS}\ge\mcrit$ binaries.

In our simulations, we consider both pair instability and pulsation pair instability supernovae for BH formation\citep{bel2016b} and include models where BHs and NSs are born as a result of rapid and delayed supernova \citep[SN;][]{fryer2012}, the maximum possible NS mass being of about $2.5\msun$. The amount and fraction of the supernova material fallback are provided by the chosen remnant-mass scheme. The remnant natal kick is slowed down based on the fallback fraction. Without any fallback modulation, we take the remnant natal kick to follow a Maxwellian distribution with velocity dispersion $\sigma \sim 265\kms$, based on observed kick distribution of Galactic NSs \citep{hobbs2005}. However, NSs that are products of the electron-capture supernova (ECS) are assumed to have a natal kick of the order of $\sim 5\kms$ \citep{pod2004}. The fallback-modulated BH natal kicks are assigned either assuming momentum conservation \citep{fryerkalo2001} or collapse asymmetry \citep{burrows1996,fryer2004}. For what concerns the BH natal spins, we consider two different models, where the prescriptions of the Geneva stellar evolution code \citep{EggenbergerMeynet2008,EkstromGeorgy2012} and \textsc{mesa} stellar evolution code \citep{PaxtonBildsten2011,PaxtonMarchant2015} are used, respectively \citep{Belczynski2020}. We also run models where the initial spin of BHs is assumed to be $0.01$, consistent with the findings of \citet{FullerMa2019}. For full details see \citet{baner2019bse} and \citet{Banerjee2021}.

We note that the model cluster set utilized in the present work is a set in progress and has limitations, which are discussed in detail in \citet{Banerjee2021} (the paper's Sec.~4 \& 5), \citet{Banerjee_2020} (the paper's Sec.~II. A.), and \citet{Banerjee2021b} (the paper's Sec.~4). In particular, mathematically rather simplistic, smooth, and spherically symmetric Plummer initial profiles are used. Therefore, alternative, more elaborate model-cluster initial conditions that are often used in the literature such as the King profile \citep[e.g.,][]{Giersz_2019,Rizzuto2021} and fractal (clumpy) initial conditions \citep[e.g.,][]{DiCarloMapelli2020} should be explored in the future simulations. We note, however, that as discussed in \citet{Banerjee_2020}, a specific choice of the initial profile is unlikely to largely influence the GR-merger outcomes from the models.

We evolve all models until $11$ Gyr, unless the cluster is dissolved earlier.

\section{Comparison to LIGO/Virgo catalogs}
\label{sect:res}

\begin{figure} 
\centering
\includegraphics[scale=0.565]{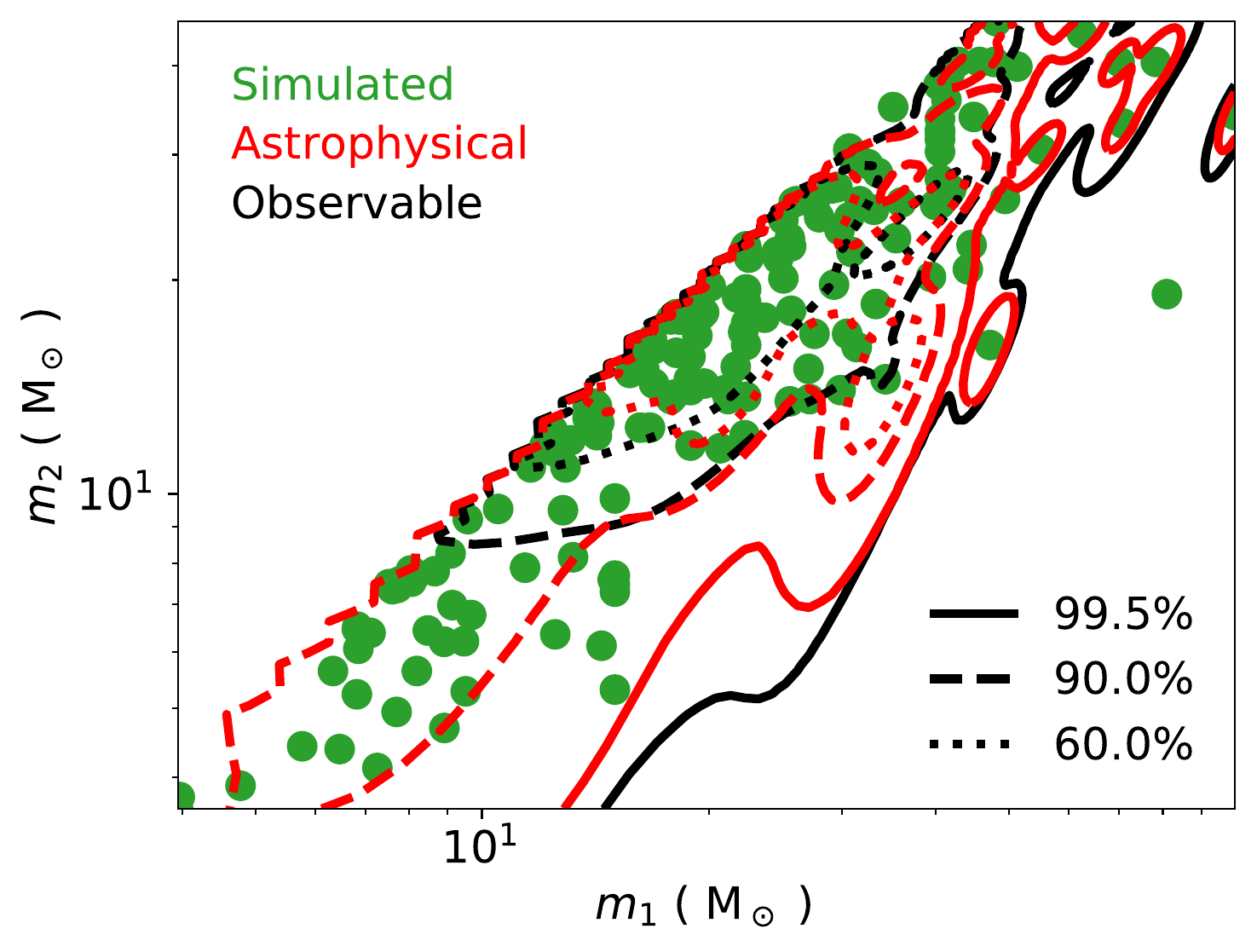}
\caption{Component masses ($m_1$ and $m_2\le m_1$) of the simulated BBH population that we extract from the cluster models (green circles) and the contour plots ($60\%$, $90\%$, $99.5\%$) of the reconstructed astrophysical and observable population.}
\label{fig:m1m2scatt}
\end{figure}

We extract the population of $195$ merging BBHs from the 65 $N$-body models described in the previous Section. As discussed in \citet{Banerjee2021}, the majority of the BBH mergers from these models are in-cluster, dynamical BBH mergers. To translate our simulated population into an observable population, we resample the model BBH mergers by accounting for both a cosmologically-motivated model for the formation of young massive and open clusters (astrophysical population) and the detection probability of LIGO/Virgo (observable population), using weights $w_{\rm cl}$ and $w_{\rm det}$, respectively. We detail our procedure in what follows.

To place the dynamically-formed BBHs in a cosmological context, we assign to each cluster a formation time $t_{\rm form}$ by sampling the cluster formation  redshift $z_{\rm form}$ from the cosmic star formation history of \citet{MadauDickinson2014}
\begin{equation}
\Psi(z)=0.01 \frac{(1+z)^{2.6}} {1.0 + [(1.0 + z) / 3.2]^{6.2}}\,{\rm M}_\sun\,{\rm yr}^{-1}\,{\rm Mpc}^{-3}\,,
\label{eqn:madau}
\end{equation}
For each BBH merger, we convolve the merger (delay) time $t_{\rm delay}$ of the BBH with the distribution of formation times for clusters by drawing $100$ random cluster formation times from Eq.~\ref{eqn:madau} for that BBH. Therefore, the merger time of a BBH in our population is the cosmic time when the parent cluster formed plus the merger delay time, $t_{\rm merger}=t_{\rm form}+t_{\rm delay}$. BBHs that merge later than the present day are discarded from our analysis. Therefore, BBHs with longer delay times are more likely to be discarded in our analysis. Each BBH that is not discarded is then assigned a weight $w_{\rm cl}$ that accounts for the parent cluster's mass and metallicity. In particular, $w_{\rm cl}$ accounts both for the cluster initial mass function, which we assume of the form \citep{PortegiesZwartMcMillan2010}
\begin{equation}
f(M) \propto \frac{1}{M^2}    
\end{equation}
and the metallicity distribution at a given redshift, which we assume is described by a log-normal distribution, with mean given by \citep{MadauFragos2017}
\begin{equation}
\log \langle Z/{\rm Z}_\odot \rangle = 0.153 - 0.074 z^{1.34}
\end{equation}
and a standard deviation of 0.5 dex \citep{DvorkinSilk2015}. Thus, the weight, $w_{\rm cl}$, assigned to each BBH that merges by present day is the product of the mass and metallicity weights for its parent star cluster. This weighting procedure provides us with the underlying astrophysical distribution of sources at a given redshift interval per comoving volume.

\begin{figure} 
\centering
\includegraphics[scale=0.565]{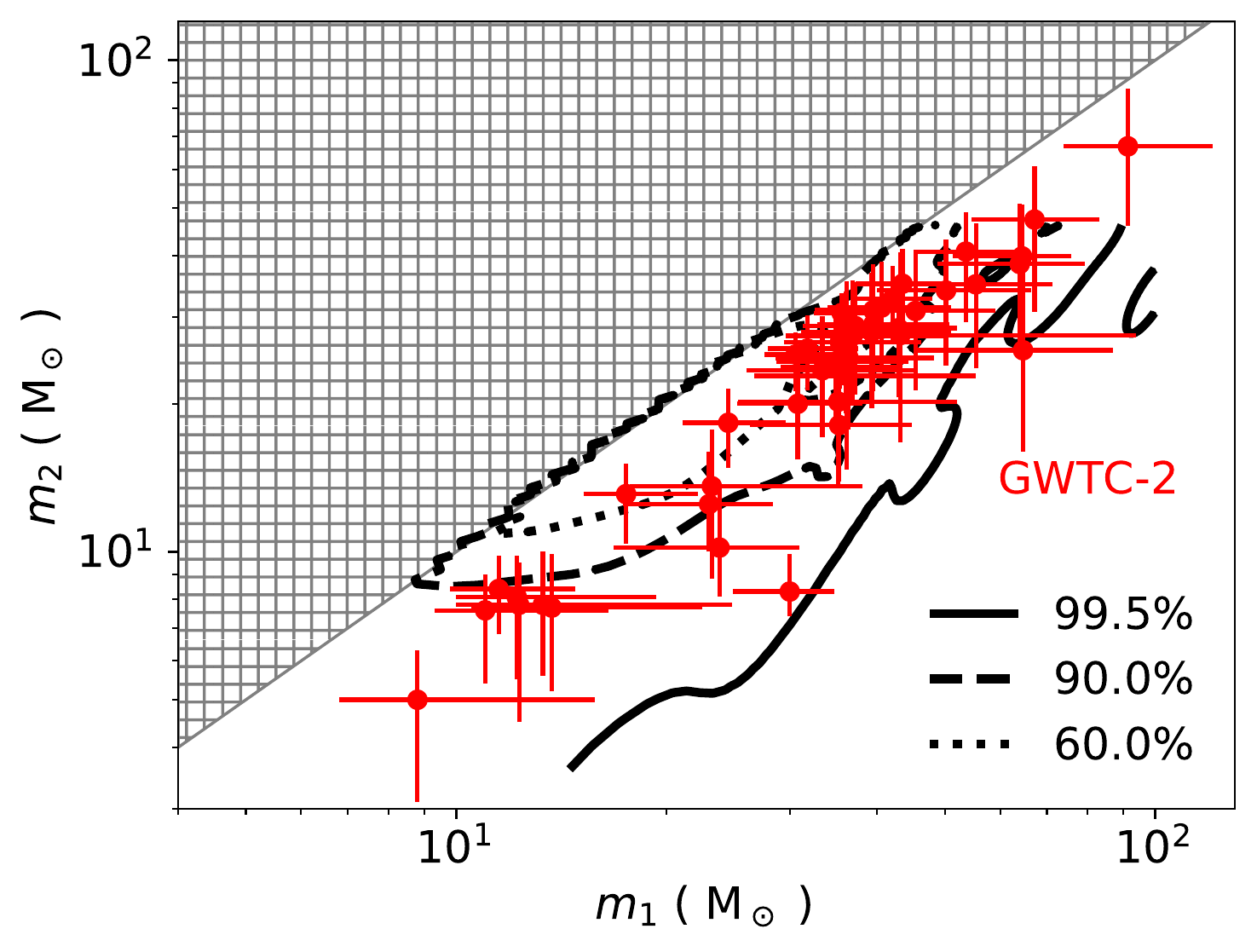}
\caption{Component masses ($m_1$ and $m_2\le m_1$) of the observable BBH population from young massive and open clusters. The contours represent the observable BBH merger population from our model. Orange circles represent observed LIGO/Virgo events and their relative error bars \citep{lvc2020cat}.}
\label{fig:m1m2}
\end{figure}

\begin{figure} 
\centering
\includegraphics[scale=0.565]{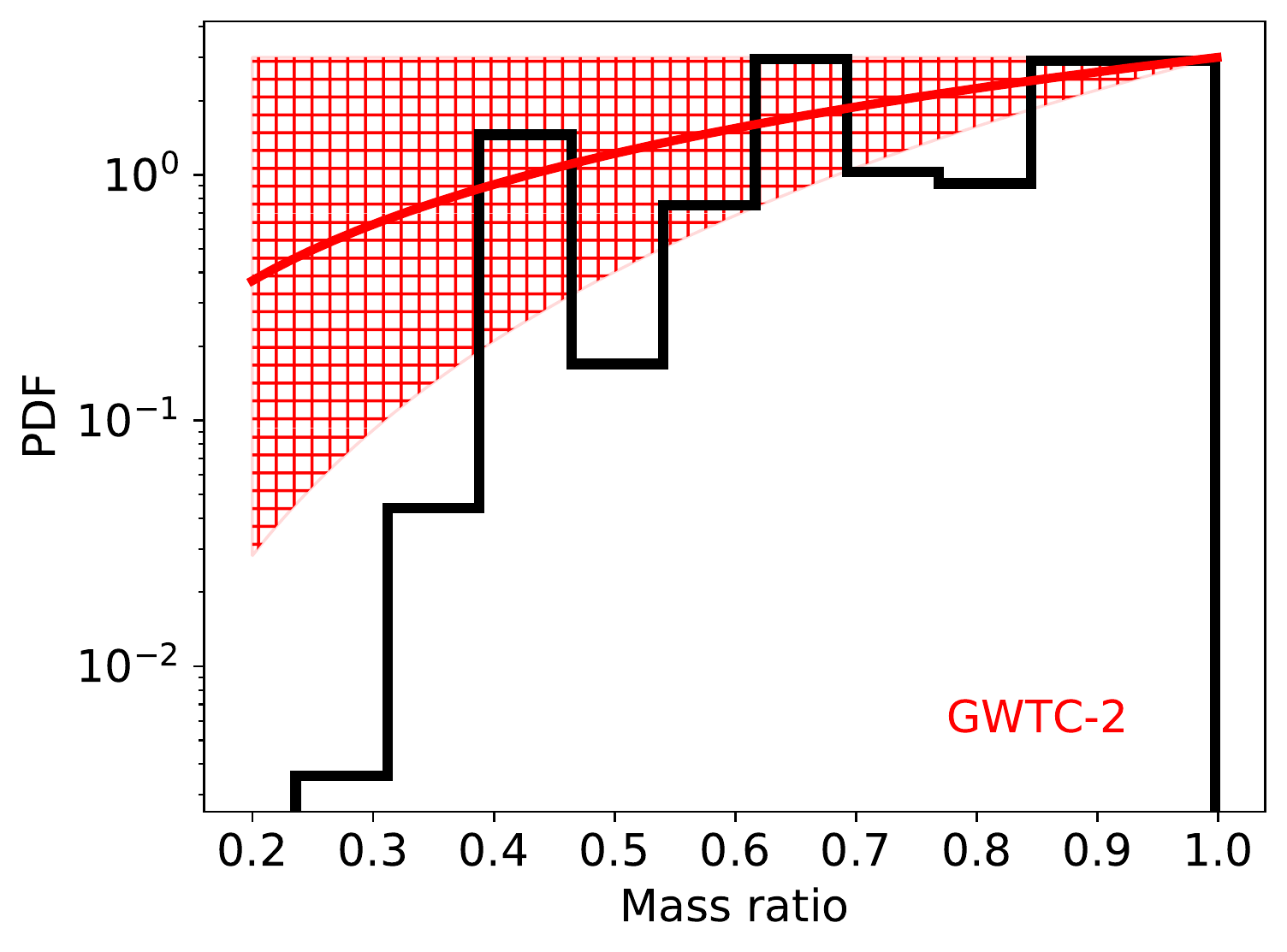}
\caption{Astrophysical mass-ratio distribution (black line) of BBH mergers from young massive and open clusters. The shaded area represent the power-law model with slope $\beta = 1.3^{+1.6}_{-1.3}$ for mass ratios inferred from GWTC-2 \citep{lvc2020catb}.}
\label{fig:mratio}
\end{figure}

In addition to populations sampled and weighted by observations of cluster formation rates, mass, and metallicity, we account for the observational weights by advanced GW observatories. Indeed, we have to take into account both the increased sensitivity of the detectors to BBHs of higher masses and the larger amount of comoving volume surveyed at higher redshifts. Therefore, we assign each BBH a detectability weight defined as \cite[see, e.g.,][]{Rodriguez2019}
\begin{equation}
w_{\rm det} = p_{\rm det}(m_1,m_2,z) \frac{dVc}{dz} \frac{dt_s}{dt_o}\,,
\label{eqn:gwweight}
\end{equation}
where $dV_c/dz$ is the amount of co-moving volume in a slice of the universe at redshift $z$, $dt_s/dt_o = 1/(1+z)$ is the difference in co-moving time between the merger redshift and the observer at $z=0$, and $p_{\rm det}(m_1,m_2,z)$ is the detection probability of sources with masses $m_1$ and $m_2$ merging at redshift $z$ that are detectable. To compute GW detectability signal-to-noise (S/N) ratio, we use the \textsc{IMRPhenomC} GW approximant \citep{SantamariaOhme2010} and assume a single LIGO instrument at design sensitivity \citep{AbbottAbbott2018lrr}, performed using \textsc{pycbc} \citep{UsmanNitz2016}. We define the detection probability $p_{\rm det}(m_1,m_2,z)$ as the fraction of sources of a given mass located at the given redshift that exceeds the detectability threshold in S/N, assuming that sources are uniformly distributed in sky location and orbital orientation \citep[e.g.][]{DominikBerti2015}
\begin{equation}
p_{\rm det}(m_1,m_2,z)=P(\rho_{\rm thr}/\rho_{\rm opt})\,,
\label{eqn:detec}
\end{equation}
where $\rho_{\rm opt}$ is the S/N ratio for an optimally located and oriented (face-on and directly overhead) binary and $\rho_{\rm thr}$ is the S/N ratio threshold, which we fix to $\rho_{\rm thr}=8$. A good approximation is given by Eq.~12 in \citet{DominikBerti2015}
\begin{eqnarray}
P(\mathcal{W})&=&a_2(1-\mathcal{W}/\alpha)^2+a_4(1-\mathcal{W}/\alpha)^4\nonumber\\
&+&a_8(1-\mathcal{W}/\alpha)^8+(1-a_2-a_4-a_8)(1-\mathcal{W}/\alpha)^{10}\,,
\end{eqnarray}
where $a_2=0.374222$, $a_4=2.04216$, $a_8=-2.63948$, and $\alpha=1.0$. 

Figure~\ref{fig:m1m2scatt} shows the distribution of the component masses ($m_1$ and $m_2<m_1$) of the simulated BBH population that we extract from the cluster models (green circles) and the contour plots ($60\%$, $90\%$, $99.5\%$) of the astrophysical and observable population, weighted using $w_{\rm cl}$ and the product of the astrophysical weights and the GW detection weights $w=w_{\rm cl}\times w_{\rm det}$, respectively. The main effect of the detection weights is to favor more massive BBH mergers, which can be detected more easily by current observatories.

\subsection{Mass distribution}

In Figure~\ref{fig:m1m2}, we show in the contours the observable primary and secondary masses, reconstructed using the mergers from our computed models. We also plot the observed LIGO/Virgo GWTC-2 events and their relative error bars \citep{lvc2020cat}. The distribution of BBH masses from young massive and open clusters agrees well with LIGO/Virgo events. As seen in Figure~\ref{fig:m1m2}, the $90\%$-confidence contour encompasses most of the GWTC-2 data points.

We note that our models rarely reproduce GW190521, a BBH merger of total mass $\sim 150\msun$, consistent with the merger of two BHs with masses of $91.4^{+29.3}_{-17.5} \msun$ and $66.9^{+15.5}_{-9.2} \msun$ \citep{ligo2020new1,ligo2020new2}. Current stellar models predict a dearth of BHs with masses larger than about $50\msun$, as a results of pulsational pair-instability process \citep{heger2003}. A likely more efficient way to produce GW190521-like events is through repeated mergers in massive and dense clusters \citep[e.g.,][]{AntoniniGieles2019,FragioneLoebRasio2020,FragioneSilk2020,MapelliSantoliquido2020,Rizzuto2021}. Thus, this process in unlikely to take place in small- and medium-mass clusters since the recoil kick imparted to the BBH merger remnant usually exceeds the cluster escape speed \citep{FragioneLoebR2020}. Alternatively, PopIII stars can produce mergers in the mass gap \citep[e.g.,][]{TanikawaSusa2021}.

\subsection{Mass-ratio distribution}

The dynamical formation of BBHs in star clusters typically involves the most massive BHs available at any given time. Even in the case where BBHs have a low mass ratio, repeated encounters preferentially exchange the lighter member of the binary and tend to create a nearly equal-mass system. As a result, the BBHs that merge in the cluster typically have nearly equal mass components drawn from the most massive BHs in the cluster.

In Figure~\ref{fig:mratio}, we show the astrophysical mass-ratio distribution of merging BBHs. We also plot the power-law model with slope $\beta = 1.3^{+1.6}_{-1.3}$ for mass ratios inferred by the LIGO/Virgo Collaboration \citep[model \textsc{POWER LAW + PEAK;}][]{lvc2020catb}. We find that the mass-ratio distribution of merging BBHs from young massive and open clusters is consistent with the inferred mass-ratio distribution from GWTC-2 within the given error bars \citep[see also][]{Banerjee2021b}. Future detections will help constrain the BBH mass-ratio distribution.

\subsection{Spin distribution}

\begin{figure} 
\centering
\includegraphics[scale=0.565]{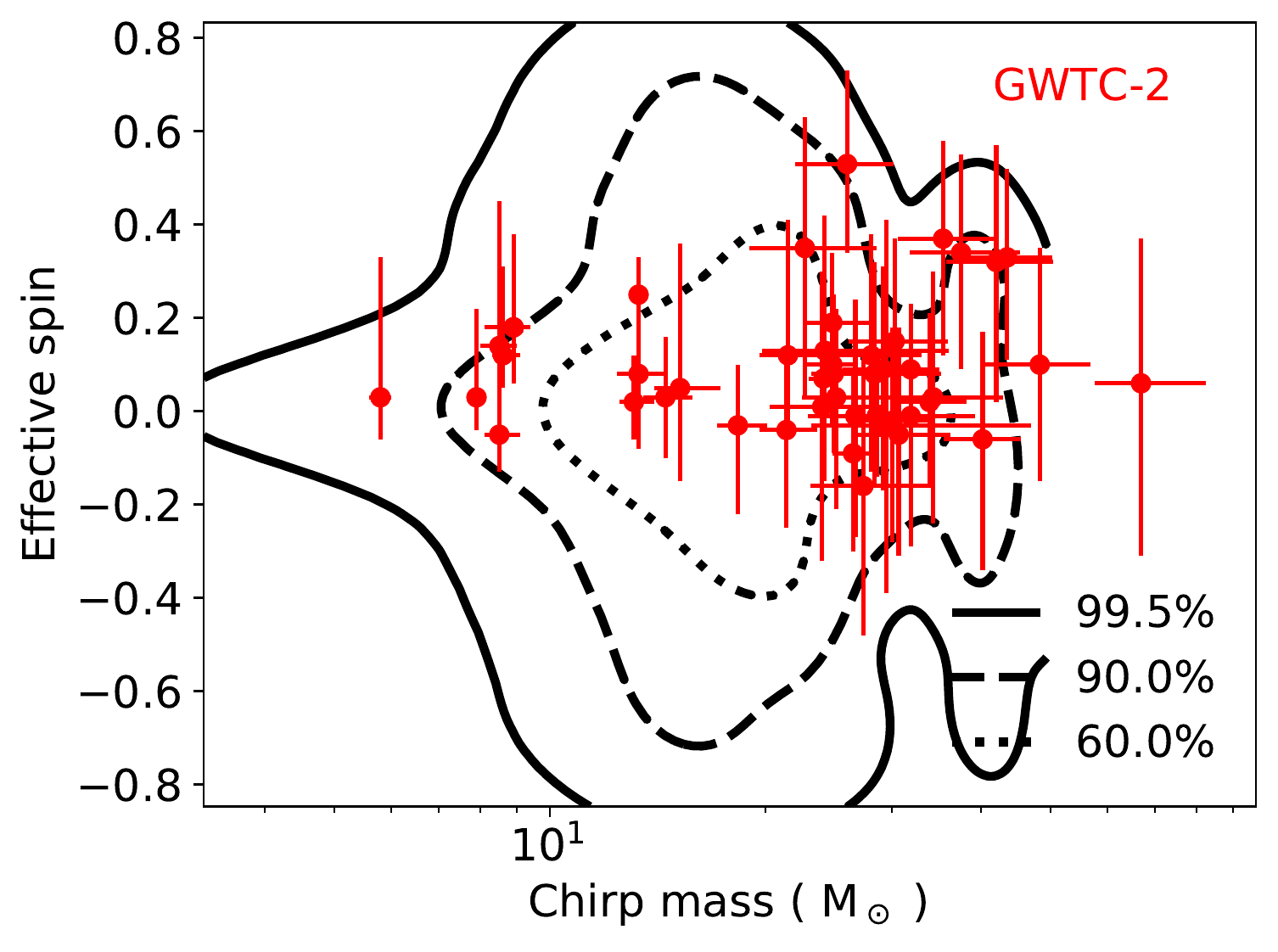}
\caption{Effective spin and chirp mass of the observable BBH population from young massive and open clusters (contours). Orange circles represent observed LIGO/Virgo events and their relative error bars \citep{lvc2020cat}.}
\label{fig:chieffmchirp}
\end{figure}

\begin{figure} 
\centering
\includegraphics[scale=0.565]{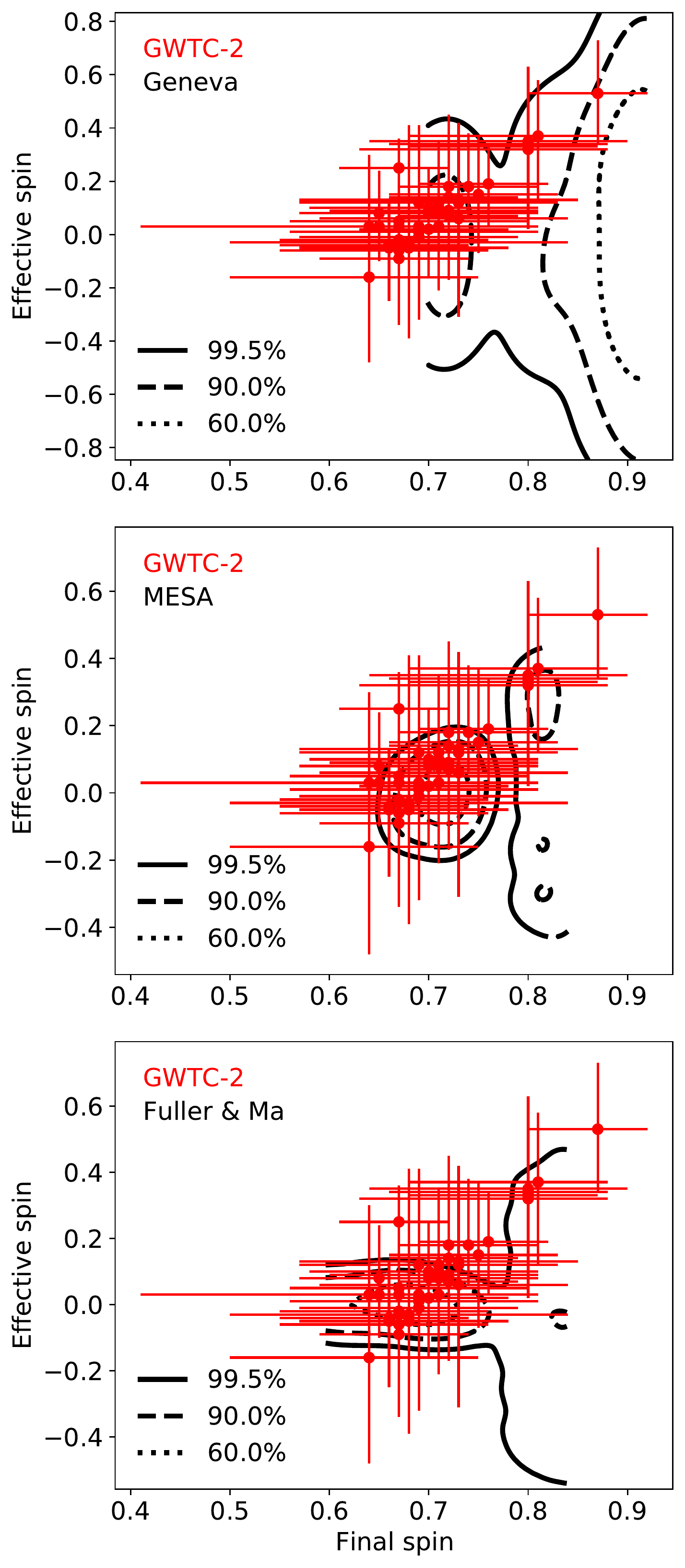}
\caption{Effective spin and final spin of the merger remnant from the observable BBH population from young massive and open clusters (contours). Orange circles represent observed LIGO/Virgo events and their relative error bars \citep{lvc2020cat}. Top: Geneva model; center: MESA model; bottom: \citet{FullerMa2019} model.}
\label{fig:chieffchifin}
\end{figure}

In Figure~\ref{fig:chieffmchirp}, we show in the contours the effective spin and chirp mass from our observable BBH population. We do discriminate among our different spin models. We also plot the observed LIGO/Virgo events and their relative error bars \citep{lvc2020cat}. We find that our distributions from young massive and open clusters agree well (within $90\%$ confidence) with LIGO/Virgo events. 

We break down the spin models in Figure~\ref{fig:chieffchifin}, where the contours show the effective spin and final spin of the merger remnant from the observable BBH population from young massive and open clusters, for the Geneva model (top), MESA model (center), and \citet{FullerMa2019} model (see \citealt{Banerjee2021} and references therein for the details of these BBH natal spin models). We compute the spin of the merger remnant using the prescriptions of \citet{Jimenez-FortezaKeitel2017}. We also plot the observed LIGO/Virgo events and their relative error bars \citep{lvc2020cat}. We find that the $99.5\%$ likelihood region of the Geneva model can explain the observed events, unlike the MESA and \citet{FullerMa2019} models. Therefore, we conclude that the Geneva model is the spin model that is the most consistent with LIGO/Virgo detected population.

Note that spins of the BHs are not taken into account in the PN evolution of the mergers. However, as discussed in \citet{Banerjee_2020,Banerjee2021}, most in-spirals begin at low frequencies ($\sim$ mHz) where the PN spin terms have very small effect on the orbital decay, so that inclusion of the spin terms would practically not alter the distribution of delay times \citep[e.g.,][]{Yu_2020}. The final in-spiral, and hence the merger configuration, within the LIGO frequency band could be influenced by the merging BHs' spins which, therefore, deserves improved treatment in the future \citep[e.g.,][]{Gerosa_2018}. Moreover, note that we have calculated the final spins separately, using updated NR fitting formulae, assuming random orientation of the BH spins, as appropriate to dynamically assembled BBHs. Despite the possible spin-orbit coupling during the final inspiral of a given BBH \citep{Antonini_2018}, this randomness in the population is not erased even just before the merger, so that the overall distribution of the final spin would remain unaffected with the inclusion of spin-orbit coupling \citep{Yu_2020}.

\section{Conclusions}
\label{sect:conc}

In this paper, we used the results of a suite of direct, high-precision $N$-body evolutionary models of young massive and open clusters to understand the detectable properties of BBHs. We have extracted the population of merging BBHs from our models and have translated it into an observable population by accounting for both a cosmologically-motivated model for the formation of young massive and open clusters and the detection probability of LIGO/Virgo. The main effect of the detection weights is to favor more massive BBH mergers, which can be detected more easily by current observatories.

We have shown that the distribution of BBH masses from young massive and open clusters agrees well with LIGO/Virgo events. However, our models rarely reproduce GW190521, which can be explained by repeated mergers in more massive clusters \citep{FragioneLoebR2020}. We have also found that the mass-ratio distribution of merging BBHs from young massive and open clusters is consistent with the inferred mass-ratio distribution from GWTC-2 within the given error bars. Finally, we have demonstrated that also the distribution of spins from the model observable BBH population are consistent with the BBH population in the GWTC-2.

With the improving sensitivity of LIGO/Virgo and the expected commissioning of KAGRA and LIGO India, hundreds of detections of merging systems are expected within the decade. Future detections will help constrain the properties of BBHs and will statistically disentangle among the contributions of the several scenarios.

\section*{Acknowledgements}

We thank the referee for their comments and suggestions which have helped improving the manuscript. GF acknowledges support from CIERA at Northwestern University. SB acknowledges the support from the Deutsche Forschungsgemeinschaft (DFG; German Research Foundation) through the individual research grant ``The dynamics of stellar mass black holes in dense stellar systems and their role in gravitational-wave generation'' (BA 4281/6-1; PI: S. Banerjee). SB acknowledges the generous support and efficient system maintenance of the computing teams at the AIfA and HISKP.\\
\ \\
Our code \textsc{gwobs} will be shared on
reasonable request to the corresponding author. Calculations of $w_{\rm det}$ are performed using \textsc{pycbc} \citep{UsmanNitz2016} and
\textsc{lal} (\url{https://git.ligo.org/lscsoft/lalsuite}).

\bibliographystyle{yahapj}
\bibliography{refs}

\begin{thebibliography}{}
\providecommand\natexlab[1]{#1}
\providecommand\JournalTitle[1]{#1}

\bibitem[{{Aarseth}(2003)}]{aseth2003}
{Aarseth}, S.~J. 2003, {Gravitational N-Body Simulations}, 430

\bibitem[{{Aarseth}(2012)}]{aseth2012}
---. 2012,
  \href{http://dx.doi.org/10.1111/j.1365-2966.2012.20666.x}{\JournalTitle{\mnras},
  422, 841}

\bibitem[{{Abbott} {et~al.}(2018){Abbott}, {Abbott}, {Abbott}, {Abernathy}, \&
  et~al.}]{AbbottAbbott2018lrr}
{Abbott}, B.~P., {Abbott}, R., {Abbott}, T.~D., {Abernathy}, M.~R., \& et~al.
  2018, \href{http://dx.doi.org/10.1007/s41114-018-0012-9}{\JournalTitle{Living
  Reviews in Relativity}, 21, 3}

\bibitem[{{Abbott} {et~al.}(2019)}]{lvc2019cat}
{Abbott}, B.~P., {et~al.} 2019,
  \href{http://dx.doi.org/10.1103/PhysRevX.9.031040}{\JournalTitle{Physical
  Review X}, 9, 031040}

\bibitem[{{Abbott} {et~al.}(2020{\natexlab{a}})}]{lvc2020cat}
{Abbott}, R., {et~al.} 2020{\natexlab{a}}, \JournalTitle{arXiv e-prints},
  arXiv:2010.14527

\bibitem[{{Abbott} {et~al.}(2020{\natexlab{b}})}]{lvc2020catb}
---. 2020{\natexlab{b}}, \JournalTitle{arXiv e-prints}, arXiv:2010.14533

\bibitem[{{Abbott} {et~al.}(2020{\natexlab{c}})}]{lvc2020catc}
---. 2020{\natexlab{c}}, \JournalTitle{arXiv e-prints}, arXiv:2010.14529

\bibitem[{{Antonini} {et~al.}(2019){Antonini}, {Gieles}, \&
  {Gualandris}}]{AntoniniGieles2019}
{Antonini}, F., {Gieles}, M., \& {Gualandris}, A. 2019,
  \href{http://dx.doi.org/10.1093/mnras/stz1149}{\JournalTitle{\mnras}, 486,
  5008}

\bibitem[{{Antonini} \& {Perets}(2012)}]{antoper12}
{Antonini}, F., \& {Perets}, H.~B. 2012,
  \href{http://dx.doi.org/10.1088/0004-637X/757/1/27}{\JournalTitle{\apj}, 757,
  27}

\bibitem[{{Antonini} {et~al.}(2018){Antonini}, {Rodriguez}, {Petrovich}, \&
  {Fischer}}]{Antonini_2018}
{Antonini}, F., {Rodriguez}, C.~L., {Petrovich}, C., \& {Fischer}, C.~L. 2018,
  \href{http://dx.doi.org/10.1093/mnrasl/sly126}{\JournalTitle{\mnras}, 480,
  L58}

\bibitem[{{Antonini} {et~al.}(2017){Antonini}, {Toonen}, \& {Hamers}}]{ant17}
{Antonini}, F., {Toonen}, S., \& {Hamers}, A.~S. 2017,
  \href{http://dx.doi.org/10.3847/1538-4357/aa6f5e}{\JournalTitle{\apj}, 841,
  77}

\bibitem[{{Askar} {et~al.}(2017){Askar}, {Szkudlarek}, {Gondek-Rosi\'{n}ska},
  {Giersz}, \& {Bulik}}]{askar17}
{Askar}, A., {Szkudlarek}, M., {Gondek-Rosi\'{n}ska}, D., {Giersz}, M., \&
  {Bulik}, T. 2017,
  \href{http://dx.doi.org/10.1093/mnrasl/slw177}{\JournalTitle{\mnras}, 464,
  L36}

\bibitem[{{Banerjee}(2018)}]{baner18}
{Banerjee}, S. 2018,
  \href{http://dx.doi.org/10.1093/mnras/stx2347}{\JournalTitle{\mnras}, 473,
  909}

\bibitem[{{Banerjee}(2020)}]{Banerjee_2020}
---. 2020,
  \href{http://dx.doi.org/10.1103/PhysRevD.102.103002}{\JournalTitle{\prd},
  102, 103002}

\bibitem[{{Banerjee}(2021{\natexlab{a}})}]{Banerjee2021}
---. 2021{\natexlab{a}},
  \href{http://dx.doi.org/10.1093/mnras/staa2392}{\JournalTitle{\mnras}, 500,
  3002}

\bibitem[{{Banerjee}(2021{\natexlab{b}})}]{Banerjee2021b}
---. 2021{\natexlab{b}},
  \href{http://dx.doi.org/10.1093/mnras/stab591}{\JournalTitle{\mnras}, 503,
  3371}

\bibitem[{{Banerjee} {et~al.}(2010){Banerjee}, {Baumgardt}, \&
  {Kroupa}}]{BanerjeeBaumgardt2010}
{Banerjee}, S., {Baumgardt}, H., \& {Kroupa}, P. 2010,
  \href{http://dx.doi.org/10.1111/j.1365-2966.2009.15880.x}{\JournalTitle{\mnras},
  402, 371}

\bibitem[{{Banerjee} {et~al.}(2020){Banerjee}, {Belczynski}, {Fryer},
  {Berczik}, {Hurley}, {Spurzem}, \& {Wang}}]{baner2019bse}
{Banerjee}, S., {Belczynski}, K., {Fryer}, C.~L., {et~al.} 2020,
  \href{http://dx.doi.org/10.1051/0004-6361/201935332}{\JournalTitle{\aap},
  639, A41}

\bibitem[{{Banerjee} \& {Kroupa}(2017)}]{banerjkroupa2017}
{Banerjee}, S., \& {Kroupa}, P. 2017,
  \href{http://dx.doi.org/10.1051/0004-6361/201526928}{\JournalTitle{\aap},
  597, A28}

\bibitem[{{Banerjee} \& {Kroupa}(2018)}]{bankroupa2018}
---. 2018, Astrophysics and Space Science Library, Vol. 424, {Formation of Very
  Young Massive Clusters and Implications for Globular Clusters}, ed.
  S.~{Stahler}, 143

\bibitem[{{Bartos} {et~al.}(2017){Bartos}, {Kocsis}, {Haiman}, \&
  {M\'{a}rka}}]{bart17}
{Bartos}, I., {Kocsis}, B., {Haiman}, Z., \& {M\'{a}rka}, S. 2017,
  \href{http://dx.doi.org/10.3847/1538-4357/835/2/165}{\JournalTitle{\apj},
  835, 165}

\bibitem[{{Belczynski} {et~al.}(2016{\natexlab{a}}){Belczynski}, {Holz},
  {Bulik}, \& {O'Shaughnessy}}]{bel16b}
{Belczynski}, K., {Holz}, D.~E., {Bulik}, T., \& {O'Shaughnessy}, R.
  2016{\natexlab{a}},
  \href{http://dx.doi.org/10.1038/nature18322}{\JournalTitle{\nat}, 534, 512}

\bibitem[{{Belczynski} {et~al.}(2016{\natexlab{b}}){Belczynski}, {Heger},
  {Gladysz}, {Ruiter}, {Woosley}, {Wiktorowicz}, {Chen}, {Bulik},
  {O'Shaughnessy}, {Holz}, {Fryer}, \& {Berti}}]{bel2016b}
{Belczynski}, K., {Heger}, A., {Gladysz}, W., {et~al.} 2016{\natexlab{b}},
  \href{http://dx.doi.org/10.1051/0004-6361/201628980}{\JournalTitle{\aap},
  594, A97}

\bibitem[{{Belczynski} {et~al.}(2020){Belczynski}, {Klencki}, {Fields},
  {Olejak}, {Berti}, {Meynet}, {Fryer}, {Holz}, {O'Shaughnessy}, {Brown},
  {Bulik}, {Leung}, {Nomoto}, {Madau}, {Hirschi}, {Kaiser}, {Jones}, {Mondal},
  {Chruslinska}, {Drozda}, {Gerosa}, {Doctor}, {Giersz}, {Ekstrom}, {Georgy},
  {Askar}, {Baibhav}, {Wysocki}, {Natan}, {Farr}, {Wiktorowicz}, {Coleman
  Miller}, {Farr}, \& {Lasota}}]{Belczynski2020}
{Belczynski}, K., {Klencki}, J., {Fields}, C.~E., {et~al.} 2020,
  \href{http://dx.doi.org/10.1051/0004-6361/201936528}{\JournalTitle{\aap},
  636, A104}

\bibitem[{{Bouffanais} {et~al.}(2021){Bouffanais}, {Mapelli}, {Santoliquido},
  {Giacobbo}, \& et~al.}]{BouffanaisMapelli2021}
{Bouffanais}, Y., {Mapelli}, M., {Santoliquido}, F., {Giacobbo}, N., \& et~al.
  2021, \JournalTitle{arXiv e-prints}, arXiv:2102.12495

\bibitem[{{Burrows} \& {Hayes}(1996)}]{burrows1996}
{Burrows}, A., \& {Hayes}, J. 1996,
  \href{http://dx.doi.org/10.1103/PhysRevLett.76.352}{\JournalTitle{\prl}, 76,
  352}

\bibitem[{{de Mink} \& {Mandel}(2016)}]{demink2016}
{de Mink}, S.~E., \& {Mandel}, I. 2016,
  \href{http://dx.doi.org/10.1093/mnras/stw1219}{\JournalTitle{\mnras}, 460,
  3545}

\bibitem[{{Di Carlo} {et~al.}(2020){Di Carlo}, {Mapelli}, {Giacobbo}, {Spera},
  \& et~al.}]{DiCarloMapelli2020}
{Di Carlo}, U.~N., {Mapelli}, M., {Giacobbo}, N., {Spera}, M., \& et~al. 2020,
  \href{http://dx.doi.org/10.1093/mnras/staa2286}{\JournalTitle{\mnras}, 498,
  495}

\bibitem[{{Dominik} {et~al.}(2015){Dominik}, {Berti}, {O'Shaughnessy},
  {Mandel}, \& et~al.}]{DominikBerti2015}
{Dominik}, M., {Berti}, E., {O'Shaughnessy}, R., {Mandel}, I., \& et~al. 2015,
  \href{http://dx.doi.org/10.1088/0004-637X/806/2/263}{\JournalTitle{\apj},
  806, 263}

\bibitem[{{Duquennoy} \& {Mayor}(1991)}]{duq1991}
{Duquennoy}, A., \& {Mayor}, M. 1991, \JournalTitle{\aap}, 500, 337

\bibitem[{{Dvorkin} {et~al.}(2015){Dvorkin}, {Silk}, {Vangioni}, {Petitjean},
  \& et~al.}]{DvorkinSilk2015}
{Dvorkin}, I., {Silk}, J., {Vangioni}, E., {Petitjean}, P., \& et~al. 2015,
  \href{http://dx.doi.org/10.1093/mnrasl/slv085}{\JournalTitle{\mnras}, 452,
  L36}

\bibitem[{{Eggenberger} {et~al.}(2008){Eggenberger}, {Meynet}, {Maeder},
  {Hirschi}, \& et~al.}]{EggenbergerMeynet2008}
{Eggenberger}, P., {Meynet}, G., {Maeder}, A., {Hirschi}, R., \& et~al. 2008,
  \href{http://dx.doi.org/10.1007/s10509-007-9511-y}{\JournalTitle{\apss}, 316,
  43}

\bibitem[{{Ekstr{\"o}m} {et~al.}(2012){Ekstr{\"o}m}, {Georgy}, {Eggenberger},
  {Meynet}, \& et~al.}]{EkstromGeorgy2012}
{Ekstr{\"o}m}, S., {Georgy}, C., {Eggenberger}, P., {Meynet}, G., \& et~al.
  2012,
  \href{http://dx.doi.org/10.1051/0004-6361/201117751}{\JournalTitle{\aap},
  537, A146}

\bibitem[{{Fragione} {et~al.}(2018){Fragione}, {Grishin}, {Leigh}, {Perets}, \&
  {Perna}}]{fragrish2018}
{Fragione}, G., {Grishin}, E., {Leigh}, N.~W.~C., {Perets}, H.~B., \& {Perna},
  R. 2018, \JournalTitle{arXiv e-prints},
  \href{http://arxiv.org/abs/1811.10627}{{\sffamily arXiv:1811.10627}}

\bibitem[{{Fragione} \& {Kocsis}(2018)}]{frak18}
{Fragione}, G., \& {Kocsis}, B. 2018,
  \href{http://dx.doi.org/10.1103/PhysRevLett.121.161103}{\JournalTitle{Phys
  Rev Lett}, 121, 161103}

\bibitem[{{Fragione} \& {Kocsis}(2019)}]{fragk2019}
---. 2019, \JournalTitle{arXiv e-prints}, arXiv:1903.03112

\bibitem[{{Fragione} \& {Loeb}(2019{\natexlab{a}})}]{frl2019a}
{Fragione}, G., \& {Loeb}, A. 2019{\natexlab{a}},
  \href{http://dx.doi.org/10.1093/mnras/stz1131}{\JournalTitle{\mnras}, 486,
  4443}

\bibitem[{{Fragione} \& {Loeb}(2019{\natexlab{b}})}]{frl2019b}
---. 2019{\natexlab{b}},
  \href{http://dx.doi.org/10.1093/mnras/stz2902}{\JournalTitle{\mnras}, 490,
  4991}

\bibitem[{{Fragione} {et~al.}(2020{\natexlab{a}}){Fragione}, {Loeb}, \&
  {Rasio}}]{FragioneLoebRasio2020}
{Fragione}, G., {Loeb}, A., \& {Rasio}, F.~A. 2020{\natexlab{a}},
  \href{http://dx.doi.org/10.3847/2041-8213/ab9093}{\JournalTitle{\apjl}, 895,
  L15}

\bibitem[{{Fragione} {et~al.}(2020{\natexlab{b}}){Fragione}, {Loeb}, \&
  {Rasio}}]{FragioneLoebR2020}
---. 2020{\natexlab{b}},
  \href{http://dx.doi.org/10.3847/2041-8213/abbc0a}{\JournalTitle{\apjl}, 902,
  L26}

\bibitem[{{Fragione} \& {Silk}(2020)}]{FragioneSilk2020}
{Fragione}, G., \& {Silk}, J. 2020,
  \href{http://dx.doi.org/10.1093/mnras/staa2629}{\JournalTitle{\mnras}, 498,
  4591}

\bibitem[{{Fragione} {et~al.}(2020{\natexlab{c}}){Fragione}, {Martinez},
  {Kremer}, {Chatterjee}, {Rodriguez}, {Ye}, {Weatherford}, {Naoz}, \&
  {Rasio}}]{fragetal2020}
{Fragione}, G., {Martinez}, M. A.~S., {Kremer}, K., {et~al.}
  2020{\natexlab{c}}, \JournalTitle{arXiv e-prints}, arXiv:2007.11605

\bibitem[{{Fryer}(2004)}]{fryer2004}
{Fryer}, C.~L. 2004,
  \href{http://dx.doi.org/10.1086/382044}{\JournalTitle{\apjl}, 601, L175}

\bibitem[{{Fryer} {et~al.}(2012){Fryer}, {Belczynski}, {Wiktorowicz},
  {Dominik}, {Kalogera}, \& {Holz}}]{fryer2012}
{Fryer}, C.~L., {Belczynski}, K., {Wiktorowicz}, G., {et~al.} 2012,
  \href{http://dx.doi.org/10.1088/0004-637X/749/1/91}{\JournalTitle{\apj}, 749,
  91}

\bibitem[{{Fryer} \& {Kalogera}(2001)}]{fryerkalo2001}
{Fryer}, C.~L., \& {Kalogera}, V. 2001,
  \href{http://dx.doi.org/10.1086/321359}{\JournalTitle{\apj}, 554, 548}

\bibitem[{{Fuller} \& {Ma}(2019)}]{FullerMa2019}
{Fuller}, J., \& {Ma}, L. 2019,
  \href{http://dx.doi.org/10.3847/2041-8213/ab339b}{\JournalTitle{\apjl}, 881,
  L1}

\bibitem[{{Gerosa} {et~al.}(2018){Gerosa}, {Berti}, {O'Shaughnessy},
  {Belczynski}, {Kesden}, {Wysocki}, \& {Gladysz}}]{Gerosa_2018}
{Gerosa}, D., {Berti}, E., {O'Shaughnessy}, R., {et~al.} 2018,
  \href{http://dx.doi.org/10.1103/PhysRevD.98.084036}{\JournalTitle{\prd}, 98,
  084036}

\bibitem[{{Giacobbo} \& {Mapelli}(2018)}]{gm2018}
{Giacobbo}, N., \& {Mapelli}, M. 2018,
  \href{http://dx.doi.org/10.1093/mnras/sty1999}{\JournalTitle{\mnras}, 480,
  2011}

\bibitem[{{Giersz} {et~al.}(2019){Giersz}, {Askar}, {Wang}, {Hypki}, {Leveque},
  \& {Spurzem}}]{Giersz_2019}
{Giersz}, M., {Askar}, A., {Wang}, L., {et~al.} 2019,
  \href{http://dx.doi.org/10.1093/mnras/stz1460}{\JournalTitle{\mnras}, 487,
  2412}

\bibitem[{{Gond{\'a}n} {et~al.}(2018){Gond{\'a}n}, {Kocsis}, {Raffai}, \&
  {Frei}}]{gondan2018}
{Gond{\'a}n}, L., {Kocsis}, B., {Raffai}, P., \& {Frei}, Z. 2018,
  \href{http://dx.doi.org/10.3847/1538-4357/aabfee}{\JournalTitle{\apj}, 860,
  5}

\bibitem[{{Grishin} {et~al.}(2018){Grishin}, {Perets}, \& {Fragione}}]{grish18}
{Grishin}, E., {Perets}, H.~B., \& {Fragione}, G. 2018,
  \href{http://dx.doi.org/10.1093/mnras/sty2477}{\JournalTitle{\mnras}, 481,
  4907}

\bibitem[{{Heger} {et~al.}(2003){Heger}, {Fryer}, {Woosley}, {Langer}, \&
  {Hartmann}}]{heger2003}
{Heger}, A., {Fryer}, C.~L., {Woosley}, S.~E., {Langer}, N., \& {Hartmann},
  D.~H. 2003, \href{http://dx.doi.org/10.1086/375341}{\JournalTitle{\apj}, 591,
  288}

\bibitem[{{Hobbs} {et~al.}(2005){Hobbs}, {Lorimer}, {Lyne}, \&
  {Kramer}}]{hobbs2005}
{Hobbs}, G., {Lorimer}, D.~R., {Lyne}, A.~G., \& {Kramer}, M. 2005,
  \href{http://dx.doi.org/10.1111/j.1365-2966.2005.09087.x}{\JournalTitle{\mnras},
  360, 974}

\bibitem[{{Jim{\'e}nez-Forteza} {et~al.}(2017){Jim{\'e}nez-Forteza}, {Keitel},
  {Husa}, {Hannam}, \& et~al.}]{Jimenez-FortezaKeitel2017}
{Jim{\'e}nez-Forteza}, X., {Keitel}, D., {Husa}, S., {Hannam}, M., \& et~al.
  2017,
  \href{http://dx.doi.org/10.1103/PhysRevD.95.064024}{\JournalTitle{\prd}, 95,
  064024}

\bibitem[{{Kremer} {et~al.}(2020){Kremer}, {Ye}, {Rui}, {Weatherford},
  {Chatterjee}, {Fragione}, {Rodriguez}, {Spera}, \& {Rasio}}]{kremer2020}
{Kremer}, K., {Ye}, C.~S., {Rui}, N.~Z., {et~al.} 2020,
  \href{http://dx.doi.org/10.3847/1538-4365/ab7919}{\JournalTitle{\apjs}, 247,
  48}

\bibitem[{{Kroupa}(2001)}]{kroupa2001}
{Kroupa}, P. 2001,
  \href{http://dx.doi.org/10.1046/j.1365-8711.2001.04022.x}{\JournalTitle{\mnras},
  322, 231}

\bibitem[{{Kruckow} {et~al.}(2018){Kruckow}, {Tauris}, {Langer}, {Kramer}, \&
  {Izzard}}]{kruc2018}
{Kruckow}, M.~U., {Tauris}, T.~M., {Langer}, N., {Kramer}, M., \& {Izzard},
  R.~G. 2018,
  \href{http://dx.doi.org/10.1093/mnras/sty2190}{\JournalTitle{\mnras}, 481,
  1908}

\bibitem[{{Li} {et~al.}(2021){Li}, {Dempsey}, {Li}, {Li}, \&
  et~al.}]{LiDempsey2021}
{Li}, Y.-P., {Dempsey}, A.~M., {Li}, S., {Li}, H., \& et~al. 2021,
  \JournalTitle{arXiv e-prints}, arXiv:2101.09406

\bibitem[{{Liu} \& {Lai}(2019)}]{liu2019}
{Liu}, B., \& {Lai}, D. 2019,
  \href{http://dx.doi.org/10.1093/mnras/sty3432}{\JournalTitle{\mnras}, 483,
  4060}

\bibitem[{{Madau} \& {Dickinson}(2014)}]{MadauDickinson2014}
{Madau}, P., \& {Dickinson}, M. 2014,
  \href{http://dx.doi.org/10.1146/annurev-astro-081811-125615}{\JournalTitle{\araa},
  52, 415}

\bibitem[{{Madau} \& {Fragos}(2017)}]{MadauFragos2017}
{Madau}, P., \& {Fragos}, T. 2017,
  \href{http://dx.doi.org/10.3847/1538-4357/aa6af9}{\JournalTitle{\apj}, 840,
  39}

\bibitem[{{Mapelli} {et~al.}(2020){Mapelli}, {Santoliquido}, {Bouffanais},
  {Arca Sedda}, \& et~al.}]{MapelliSantoliquido2020}
{Mapelli}, M., {Santoliquido}, F., {Bouffanais}, Y., {Arca Sedda}, M., \&
  et~al. 2020, \JournalTitle{arXiv e-prints}, arXiv:2007.15022

\bibitem[{{Marchant} {et~al.}(2016){Marchant}, {Langer}, {Podsiadlowski},
  {Tauris}, \& {Moriya}}]{march16}
{Marchant}, P., {Langer}, N., {Podsiadlowski}, P., {Tauris}, T.~M., \&
  {Moriya}, T.~J. 2016,
  \href{http://dx.doi.org/10.1051/0004-6361/201628133}{\JournalTitle{A\& A},
  588, A50}

\bibitem[{{Michaely} \& {Perets}(2020)}]{MichaelyPerets2020}
{Michaely}, E., \& {Perets}, H.~B. 2020,
  \href{http://dx.doi.org/10.1093/mnras/staa2720}{\JournalTitle{\mnras}, 498,
  4924}

\bibitem[{{Moe} \& {Di Stefano}(2017)}]{moedist2017}
{Moe}, M., \& {Di Stefano}, R. 2017,
  \href{http://dx.doi.org/10.3847/1538-4365/aa6fb6}{\JournalTitle{\apjs}, 230,
  15}

\bibitem[{{O'Leary} {et~al.}(2009){O'Leary}, {Kocsis}, \& {Loeb}}]{olea09}
{O'Leary}, R.~M., {Kocsis}, B., \& {Loeb}, A. 2009,
  \href{http://dx.doi.org/10.1111/j.1365-2966.2009.14653.x}{\JournalTitle{\mnras},
  395, 2127}

\bibitem[{{O'Leary} {et~al.}(2016){O'Leary}, {Meiron}, \& {Kocsis}}]{olea16}
{O'Leary}, R.~M., {Meiron}, Y., \& {Kocsis}, B. 2016,
  \href{http://dx.doi.org/10.3847/2041-8205/824/1/L12}{\JournalTitle{\apj
  Lett}, 824, L12}

\bibitem[{{Paxton} {et~al.}(2011){Paxton}, {Bildsten}, {Dotter}, {Herwig}, \&
  et~al.}]{PaxtonBildsten2011}
{Paxton}, B., {Bildsten}, L., {Dotter}, A., {Herwig}, F., \& et~al. 2011,
  \href{http://dx.doi.org/10.1088/0067-0049/192/1/3}{\JournalTitle{\apjs}, 192,
  3}

\bibitem[{{Paxton} {et~al.}(2015){Paxton}, {Marchant}, {Schwab}, {Bauer}, \&
  et~al.}]{PaxtonMarchant2015}
{Paxton}, B., {Marchant}, P., {Schwab}, J., {Bauer}, E.~B., \& et~al. 2015,
  \href{http://dx.doi.org/10.1088/0067-0049/220/1/15}{\JournalTitle{\apjs},
  220, 15}

\bibitem[{{Perna} {et~al.}(2019){Perna}, {Wang}, {Farr}, {Leigh}, \&
  et~al.}]{PernaWang2019}
{Perna}, R., {Wang}, Y.-H., {Farr}, W.~M., {Leigh}, N., \& et~al. 2019,
  \href{http://dx.doi.org/10.3847/2041-8213/ab2336}{\JournalTitle{\apjl}, 878,
  L1}

\bibitem[{{Petrovich} \& {Antonini}(2017)}]{petr17}
{Petrovich}, C., \& {Antonini}, F. 2017,
  \href{http://dx.doi.org/10.3847/1538-4357/aa8628}{\JournalTitle{\apj}, 846,
  146}

\bibitem[{{Plummer}(1911)}]{plummer1911}
{Plummer}, H.~C. 1911,
  \href{http://dx.doi.org/10.1093/mnras/71.5.460}{\JournalTitle{\mnras}, 71,
  460}

\bibitem[{{Podsiadlowski} {et~al.}(2004){Podsiadlowski}, {Langer},
  {Poelarends}, {Rappaport}, {Heger}, \& {Pfahl}}]{pod2004}
{Podsiadlowski}, P., {Langer}, N., {Poelarends}, A.~J.~T., {et~al.} 2004,
  \href{http://dx.doi.org/10.1086/421713}{\JournalTitle{\apj}, 612, 1044}

\bibitem[{{Portegies Zwart} {et~al.}(2010{\natexlab{a}}){Portegies Zwart},
  {McMillan}, \& {Gieles}}]{portgz2010}
{Portegies Zwart}, S.~F., {McMillan}, S. L.~W., \& {Gieles}, M.
  2010{\natexlab{a}},
  \href{http://dx.doi.org/10.1146/annurev-astro-081309-130834}{\JournalTitle{\araa},
  48, 431}

\bibitem[{{Portegies Zwart} {et~al.}(2010{\natexlab{b}}){Portegies Zwart},
  {McMillan}, \& {Gieles}}]{PortegiesZwartMcMillan2010}
---. 2010{\natexlab{b}},
  \href{http://dx.doi.org/10.1146/annurev-astro-081309-130834}{\JournalTitle{\araa},
  48, 431}

\bibitem[{{Rasskazov} \& {Kocsis}(2019)}]{rass2019}
{Rasskazov}, A., \& {Kocsis}, B. 2019, \JournalTitle{arXiv e-prints},
  arXiv:1902.03242

\bibitem[{{Rizzuto} {et~al.}(2021)}]{Rizzuto2021}
{Rizzuto}, F.~P., {et~al.} 2021, \JournalTitle{\mnras}, 501, 5257

\bibitem[{{Rodriguez} {et~al.}(2018){Rodriguez}, {Amaro-Seoane}, {Chatterjee},
  \& {Rasio}}]{rod18}
{Rodriguez}, C.~L., {Amaro-Seoane}, P., {Chatterjee}, S., \& {Rasio}, F.~A.
  2018,
  \href{http://dx.doi.org/10.1103/PhysRevLett.120.151101}{\JournalTitle{PRL},
  120, 151101}

\bibitem[{{Rodriguez} {et~al.}(2019){Rodriguez}, {Zevin}, {Amaro-Seoane},
  {Chatterjee}, {Kremer}, {Rasio}, \& {Ye}}]{Rodriguez2019}
{Rodriguez}, C.~L., {Zevin}, M., {Amaro-Seoane}, P., {et~al.} 2019,
  \href{http://dx.doi.org/10.1103/PhysRevD.100.043027}{\JournalTitle{\prd},
  100, 043027}

\bibitem[{{Sana} \& {Evans}(2011)}]{sanaevans2011}
{Sana}, H., \& {Evans}, C.~J. 2011,
  \href{http://dx.doi.org/10.1017/S1743921311011124}{in IAU Symposium, Vol.
  272, Active OB Stars: Structure, Evolution, Mass Loss, and Critical Limits,
  ed. C.~{Neiner}, G.~{Wade}, G.~{Meynet}, \& G.~{Peters}}, 474

\bibitem[{{Santamar{\'\i}a} {et~al.}(2010){Santamar{\'\i}a}, {Ohme}, {Ajith},
  {Br{\"u}gmann}, \& et~al.}]{SantamariaOhme2010}
{Santamar{\'\i}a}, L., {Ohme}, F., {Ajith}, P., {Br{\"u}gmann}, B., \& et~al.
  2010,
  \href{http://dx.doi.org/10.1103/PhysRevD.82.064016}{\JournalTitle{\prd}, 82,
  064016}

\bibitem[{{Secunda} {et~al.}(2019){Secunda}, {Bellovary}, {Mac Low}, {Ford},
  {McKernan}, {Leigh}, {Lyra}, \& {S{\'a}ndor}}]{secunda2019}
{Secunda}, A., {Bellovary}, J., {Mac Low}, M.-M., {et~al.} 2019,
  \href{http://dx.doi.org/10.3847/1538-4357/ab20ca}{\JournalTitle{\apj}, 878,
  85}

\bibitem[{{Silsbee} \& {Tremaine}(2017)}]{sil17}
{Silsbee}, K., \& {Tremaine}, S. 2017,
  \href{http://dx.doi.org/10.3847/1538-4357/aa5729}{\JournalTitle{\apj}, 836,
  39}

\bibitem[{{Spitzer}(1987)}]{spitzer1987}
{Spitzer}, L. 1987, {Dynamical evolution of globular clusters}

\bibitem[{{Tanikawa} {et~al.}(2021){Tanikawa}, {Susa}, {Yoshida}, {Trani}, \&
  et~al.}]{TanikawaSusa2021}
{Tanikawa}, A., {Susa}, H., {Yoshida}, T., {Trani}, A.~A., \& et~al. 2021,
  \href{http://dx.doi.org/10.3847/1538-4357/abe40d}{\JournalTitle{\apj}, 910,
  30}

\bibitem[{{The LIGO Scientific Collaboration} \& {the Virgo
  Collaboration}(2020{\natexlab{a}})}]{ligo2020new1}
{The LIGO Scientific Collaboration}, \& {the Virgo Collaboration}.
  2020{\natexlab{a}}, \JournalTitle{arXiv e-prints}, arXiv:2009.01075

\bibitem[{{The LIGO Scientific Collaboration} \& {the Virgo
  Collaboration}(2020{\natexlab{b}})}]{ligo2020new2}
---. 2020{\natexlab{b}},
  \href{http://dx.doi.org/10.3847/2041-8213/aba493}{\JournalTitle{\apjl}, 900,
  L13}

\bibitem[{{Trani} {et~al.}(2021){Trani}, {Tanikawa}, {Fujii}, {Leigh}, \&
  {Kumamoto}}]{Trani2021}
{Trani}, A.~A., {Tanikawa}, A., {Fujii}, M.~S., {Leigh}, N.~W.~C., \&
  {Kumamoto}, J. 2021,
  \href{http://dx.doi.org/10.1093/mnras/stab967}{\JournalTitle{\mnras}, 504,
  910}

\bibitem[{{Usman} {et~al.}(2016){Usman}, {Nitz}, {Harry}, {Biwer}, \&
  et~al.}]{UsmanNitz2016}
{Usman}, S.~A., {Nitz}, A.~H., {Harry}, I.~W., {Biwer}, C.~M., \& et~al. 2016,
  \href{http://dx.doi.org/10.1088/0264-9381/33/21/215004}{\JournalTitle{Classical
  and Quantum Gravity}, 33, 215004}

\bibitem[{{Wong} {et~al.}(2020){Wong}, {Breivik}, {Kremer}, \&
  {Callister}}]{WongBreivik2020}
{Wong}, K. W.~K., {Breivik}, K., {Kremer}, K., \& {Callister}, T. 2020,
  \JournalTitle{arXiv e-prints}, arXiv:2011.03564

\bibitem[{{Yu} {et~al.}(2020){Yu}, {Ma}, {Giesler}, \& {Chen}}]{Yu_2020}
{Yu}, H., {Ma}, S., {Giesler}, M., \& {Chen}, Y. 2020,
  \href{http://dx.doi.org/10.1103/PhysRevD.102.123009}{\JournalTitle{\prd},
  102, 123009}

\bibitem[{{Zevin} {et~al.}(2020){Zevin}, {Bavera}, {Berry}, {Kalogera}, \&
  et~al.}]{ZevinBavera2020}
{Zevin}, M., {Bavera}, S.~S., {Berry}, C. P.~L., {Kalogera}, V., \& et~al.
  2020, \JournalTitle{arXiv e-prints}, arXiv:2011.10057

\end{thebibliography}

\end{document}